# Is QBism a Possible Solution to the Conceptual Problems of Quantum Mechanics?[1]

## Hervé Zwirn

CMLA (ENS Paris Saclay, 61 avenue du Président Wilson 94235 Cachan, France)

&

IHPST (CNRS, ENS Ulm, University Paris 1, 13 rue du Four, 75006 Paris, France)

herve.zwirn@gmail.com

**Abstract:** QBism is a recently developed version of Quantum Bayesianism. QBists think that the primitive concept of experience is the central subject of science. QBism refuses the idea that the quantum state of a system is an objective description of this system. It is a tool for assigning a subjective probability to the agent's future experience. So quantum mechanics does not directly say something about the "external world". A measurement (in the usual sense) is just a special case of what QBism calls experience and this is any action done by any agent on her external world. The result of the measurement is the experience that the agent gets from her action on her personal world. For Qbists, this way of seeing the measurement helps rejecting any non-local influences. We will examine if these claims are robust and will show that a certain number of issues needs to be clarified in order to get a coherent picture. Depending on the way these necessary clarifications are made, the global image given by Qbism can vary from pure instrumentalism to interpretations making the observer play a radical role.

## 1. INTRODUCTION

Quantum Bayesianism is a type of interpretation that applies the Bayesian interpretation of probability to Quantum Mechanics. For the Bayesian interpretation everything starts by assuming some prior probabilities. Then some recipes are provided to update these probabilities on new information. For the subjective Bayesians (de Finetti 1990, Ramsey 1931, Savage 1954), probabilities are related to an epistemic (hence personal and subjective) uncertainty and represent the degree of belief of an agent for the happening of an event.

There are several versions of Quantum Bayesianism (Pitowsky 2002, Baez 2003, Bub & Pitowsky 2010, Bub 2019, Youssef 1994, Leifer & Spekkens 2013, Caticha 2007).

QBism is the current state of the version of Quantum Bayesianism that was initially developed by Caves, Fuchs and Schack (2002, 2007) but has evolved in such a way that Caves now "*does not subscribe to it*" (Stacey, 2019b). It has been more recently adopted by Mermin (2013, 2014, 2016, Fuchs,

---

[1] This paper is to appear in the forthcoming Oxford Handbook of the History of Interpretations and Foundations of Quantum Mechanics.

Mermin & Schack 2014, 2015). The denomination QBism appeared for the first time in a 2009 arXiv article which has been published in 2013 (Fuchs & Schack 2013). As Fuchs (2016) says:

*"The term QBism, invented in 2009, initially stood for Quantum Bayesianism, a view of quantum theory a few of us had been developing since 1993. Eventually, however, I.J. Good's warning that there are 46,656 varieties of Bayesianism came to bite us, with some Bayesians feeling their good name had been hijacked. David Mermin suggested that the B in QBism should more accurately stand for "Bruno" as in Bruno de Finetti, so that we would at least get the variety of (subjective) Bayesianism right. The trouble is QBism incorporates a kind metaphysics that even Bruno de Finetti might have rejected! So, trying to be as true to our story as possible, we momentarily toyed with the idea of associating the B with what the early 20$^{th}$-century United State Supreme Court Justice Oliver Wendell Homes Jr called "bettabilitarianism". It is the idea that the world is loose at the joints, that indeterminism plays a real role in the world. […] But what an ugly, ugly word, bettabilitarianism! Therefore, maybe one should just think of the B as standing for no word in particular, but a deep idea instead: That the world is so wired that our actions as active agents actually matter."*

So, QBists think that the central subject of science is the primitive concept of experience. This seems similar to Bohr's position who insisted on saying that experiments were the basic subject of quantum mechanics, but what Bohr had in mind was the classical reading of a macroscopic apparatus, while for QBism "experience" is much more than that. The word "experience" must be understood here as "personal experience", meaning for each agent what the world has induced in her throughout the course of her life. Experience is the way in which the world impinges on any agent and how the agent impinges on the world.

QBists adopt the subjective interpretation of probability according to which probabilities represent the degrees of belief of an agent and, hence, are particular to that agent. This is an important point to notice: as they say, it is a "single user theory". This means that:

*"Probability assignments express the beliefs of the agent who makes them, and refer to that same agent's expectations for her subsequent experiences"* (Fuchs, Mermin & Schack 2014).

*"When I -the agent- write down a quantum state it is my quantum state, no one else's. When I contemplate a measurement, I contemplate its results, outcomes, consequences for me, no one else –it is my experience"* (Fuchs 2018).

According to QBism the entities of the formalism (wave functions, observables, Hamiltonians, probabilities) are not objective and attached to an external reality. They are subjective in nature and have no absolute value but are relative to a particular agent. So they are not necessarily the same for all observers. QBists refuse the idea that the quantum state of a system is an objective property of that system (Fuchs & Schack 2015). It is only a tool for assigning a subjective probability to the agent's future experience. It only represents the degree of belief of the agent about the possible outcomes of future measurements. So quantum mechanics does not directly say something about the "external world".

*"But quantum mechanics itself does not deal directly with the objective world; it deals with the experiences of that objective world that belong to whatever particular agent is making use of the quantum theory."* (Fuchs, Mermin & Schack 2014).

This means that QBism does not consider quantum mechanics as a theory describing the external world that we could gradually discover more and more intimately through our measurements, but as a theory that describes the agent's experience. A measurement (in the usual sense) is just a special case of what QBism calls experience, and that is any action taken by any agent on her external world. The result of the measurement is the experience the agent gets from her action on her personal world. In this sense, measurements are made continuously by every agent. A measurement does not reveal a pre-existing state of affairs but creates a result for the agent.

*"A measurement does not, as the term unfortunately suggests, reveal a pre-existing state of affairs. It is an action on the world by an agent that results in the creation of an outcome - a new experience for that agent."* (Fuchs, Mermin & Schack 2014).

So, QBism could be seen as an instrumentalist interpretation since it presents quantum mechanics as nothing but a tool allowing any agent to compute her probabilistic expectations for her future experience from the knowledge of the results of her past experience. Then the goal of quantum formalism is only to give recipes to allow agents to compute their personal degree of belief about what will happen if they do such and such experiment. But QBists refuse to be called instrumentalist:

*"QBism has also breathed new life into instrumentalism (a term originally coined in by John Dewey to describe his brand of pragmatism). Instrumentalism, it seems, has a bad reputation. It is perceived as unable to explain the successes of our best theories, and as unwilling to reach for nature's essence. QBism tells us that this assessment may well be premature. What QBism shows is that an enriched instrumentalist attitude— in this case, seeing quantum theory as something that "deals only with the object– subject relation," as Schrödinger once put it—may in fact be our best bet of arriving at explanations more profound and radical than anything we could have imagined."* (Schlosshauer's foreword in Fuchs 2015).

QBists claim that they endorse a special kind of realism they call "participatory realism" where reality goes further than what can be said in a third-person account (Fuchs 2016).

Perhaps a good way to introduce the basics of QBism is to quote what Fuchs (2018) says about three important tenets of this interpretation in the paper where he both acknowledges what QBism owes to Bohr and makes explicit the main divergences with the Copenhagen interpretation:

*"Along the way, we lay out three tenets of QBism in some detail: 1) The Born Rule – the foundation of what quantum theory means for QBism – is a normative statement. It is about the decision-making*

*behaviour any individual agent should strive for; it is not a descriptive "law of nature" in the usual sense. 2) All probabilities, including all quantum probabilities, are so subjective they never tell nature what to do. This includes probability-1 assignments. Quantum states thus have no "ontic hold" on the world. 3) Quantum measurement outcomes just are personal experiences for the agent gambling upon them. Particularly, quantum measurement outcomes are not, to paraphrase Bohr, instances of "irreversible amplification in devices whose design is communicable in common language suitably refined by the terminology of classical physics.""*

## 2. PERSONNAL EXPERIENCE AND SUBJECTIVE PROBABILITIES

Strongly influenced by "Wheeler's vision of a world built upon "law without law", a world where the big bang is here and the observer is a participator in the process" (Fuchs 2018), QBism proposes a picture where the agent and the external world interact in a creative way to give birth to an experience for the agent. This allows them to separate the subjective from the objective, which in turn is the way to unscramble the omelette made by Heisenberg and Bohr according to Jaynes (1990), another great influencer for QBists:

*"Our present QM formalism is not purely epistemological; it is a peculiar mixture describing in part realities of Nature, in part incomplete human information about Nature - all scrambled up by Heisenberg and Bohr into an omelette that nobody has seen how to unscramble. Yet we think that the unscrambling is a prerequisite for any further advance in basic physical theory. For, if we cannot separate the subjective and objective aspects of the formalism, we cannot know what we are talking about; it is just that simple".*

QBism then proceeds to sort what is subjective and what is objective. The quantum system is something real and independent of the agent. The quantum state represents a collection of subjective degrees of belief about it and the probabilities given by the formalism for a measurement are nothing but degrees of belief concerning the results that the agent could get if she were interacting with the quantum system. Since a measurement is any action taken by any agent on her external world and since this action creates something new for the agent, probabilities are a numerical way to express the degree of belief the agent has on such and such result and are a guide for determining the amount she would bet on a particular result.

Hence, quantum theory does not describe an objective reality existing independently of any observer and where events, that any observer can passively witness, happen by themselves. On the contrary, quantum theory is a personal guide that has the same general form and the same rules for all the agents but that each agent specifies to adapt it to her experience. Hence, quantum states, observables, Hamiltonians, density matrices etc … are not absolute concepts but tools relative to each specific agent. There is no reason for two agents to share the same quantum state when considering a quantum system, nor to have the same probability for getting a particular result during a measurement. For QBists, probabilities cannot tell Nature what to do. Gambling on events are one thing, the events of the world are another. As they say, a quantum state has no "ontic hold" on the world. They go so far as saying this even for probability 1, i.e. for an outcome that is certain, there is nothing intrinsic to the quantum system, i.e., no objectively real property of the system that guarantees this particular outcome.

"*The statement that an outcome is certain to occur is always a statement relative to a scientist's (necessarily subjective) state of belief. "It is certain" is a state of belief, not a fact.*" (Caves, Fuchs & Schack 2007).

Hence even if the probability of a 'yes' outcome is exactly 1 for the measurement of an agent that "*does not mean that the world is forbidden to give her a 'no' outcome for this measurement*" because "*My probabilities cannot tell nature what to do*" (Fuchs 2018).

We will see that this is a key point in the way QBism refuse the criterion of reality in the EPR argument.

### 3. THE MEASUREMENT PROBLEM

The measurement problem in quantum mechanics comes from the fact that there are two postulates describing the dynamics of the state of a system. The first one says that the state evolves deterministically according to the Schrödinger equation which is linear and unitary, and hence transforms a superposed state into another superposed state. The second one, the reduction postulate, says that under a measurement the wave function collapses, and the state is projected onto one of the eigenstates of the observable that is measured. So, in general, these two laws do not lead to the same results. The difficulty is that it is not possible to give a non-ambiguous description of what a measurement is. Hence, there is no clear rule to decide when each one of these two postulates should be used. The question is to explain why only one result among many possible others is obtained during a measurement. The way QBists solve the problem is very simple. First, for QBists, unitary evolution concerns the change between the belief the agent has at time t0 for the outcomes of a measurement she could perform at t0 and the belief she has at time t0 about a measurement she would perform at t1 > t0. Second, for QBists a measurement is the experience of an agent. They assume from the beginning that the direct internal awareness of her private experience is the only phenomenon accessible to an agent which she does not model with quantum mechanics, and that the agent's awareness is the result of the experiment. Hence, there is no longer any ambiguity about when to use the reduction postulate, since it does not say anything about the "real state" of the system that is measured and is nothing else than the updating of the agent's state assignment on the basis of her experience. QBists like to quote Asher Peres (1978) famous sentence: "*Unperformed experiments have no results*".

In particular, there is no measurement when there is no agent: a Stern and Gerlach apparatus cannot measure by itself the spin of a particle. QBists want to stress the fact that they depart from the standard

Copenhagen interpretation in that a measurement is not "*an interaction between classical and quantum objects occurring apart from and independently of any observer*" (Fuchs & Schack 2013). They quote here the Landau and Lifshitz formulation of the Copenhagen interpretation. To be fair, this quotation is not really faithful to Bohr's position which is more subtle than that. But it is true that even when it is taken in all its subtlety, Bohr's position is not able to solve the measurement problem in a satisfying way. Indeed it does not make any explicit resort to the observer to define a measurement while simultaneously it appeals to the use that is done (by her) of the macroscopic apparatus. For QBists an apparatus should be seen as an extension of the agent like another sense organ or a prosthesis. As such by itself it cannot create a result.

For QBism, any agent can use the quantum formalism to model any system external to herself whether they be atoms, apparatuses or even other agents. For any agent, the personal internal awareness of other agents is inaccessible and not something she can apply quantum mechanics to. But communication with other agents (verbal exchanges for example) is. This means that an agent can use a description of another agent through a superposed state encoding her probabilities for the possible answers to any question she might ask before getting a definite answer. A consequence is that it seems that reality can differ from one agent to another. But a reservations should be made about the meaning of the comparison between the perceptions of two agents since this comparison could only be done from a meta-observer point of view, which does not exist. No third person point of view is allowed.

So an important feature of QBism is the fundamental role that the observer plays in the measurement process and the relinquishment of any absolute description of the world.

### 4. DECOHERENCE

Decoherence takes the interactions between the system and its environment into account (Zeh 1970, Zurek 1981, 1982, 1991). It is usually held that it plays an important role both to solve the preferred basis problem through the selection of a "pointer basis" and to show how the reduced density matrix of the system remains very nearly diagonal in this basis for a time exceeding the age of the universe. The Schrödinger equation is applied to a system and its environment so they become entangled and the reduced density matrix of the system made by tracing off the degrees of freedom of the environment is nearly diagonal. Decoherence is not a solution to the measurement problem (Zwirn 1997, 2015, 2016) but it helps understand the classical appearance of the world (Joos, Zeh, Kiefer, Giulini, Kupsh & Stamatescu 2003). In particular it explains why macroscopic objects do not exhibit interferences which the standard view of quantum mechanics predicts we should see with quantum superpositions. But for QBists the Schrödinger equation is not a law governing the behaviour of an objective quantum state. It is only a tool for describing how the agent's epistemic state evolves. Hence decoherence plays no role in their framework:

*"For a Quantum Bayesian, the only physical process in a quantum measurement is what was previously seen as "the selection step"—i.e., the agent's action on the external world and its unpredictable consequence for her, the data that leads to a new state of belief about the system. Thus, it would seem there is no foundational place for decoherence in the Quantum Bayesian program. And this is true."* (Fuchs & Schack 2012).

It is true that the main advantage of decoherence is to explain the classical appearance of the world if we take quantum states as representing real states of the systems. But for QBists, this does not need to be explained since the wave-function is merely a probability distribution:

*""Why does the world look classical if it actually operates according to quantum mechanics?" The touted mystery is that we never "see" quantum superposition and entanglement in our everyday experience. But have you ever seen a probability distribution sitting in front of you? Probabilities in personalist Bayesianism are not the sorts of things that can be seen; they are the things that are thought. It is events that are seen."* (Fuchs & Stacey 2019).

## 5. RECONTRUCTING THE QUANTUM FORMALISM FROM PROBABILITIES

Probabilities are at the heart of QBism. So it is natural to ask:

*"If quantum theory is so closely allied with probability theory, then why is it not written in a language that starts with probability, rather than a language that ends with it? Why does quantum theory invoke the mathematical apparatus of complex amplitudes, Hilbert spaces, and linear operators?"* (Fuchs & Stacey 2019).

One goal QBists try to achieve is to reconstruct the quantum formalism from probabilities, using tools coming from the growing field of quantum information theory (Fuchs & Stacey 2019). One key ingredient is a hypothetical structure called a "symmetric informationally complete positive-operator valued measure", a SIC POVM. This is a set of d2 rank-one projection operator $\prod_i = |\varphi_i\rangle \langle \varphi_i |$ on a d-dimensional Hilbert space such that $|\langle \varphi_i | \varphi_j \rangle|^2 = \frac{1}{d+1}$ whenever i≠j. What is interesting is that the operators $H_i = \frac{1}{d} \prod_i$ are positive semi-definite and sum to the identity so they can be considered as representing a generalisation of a standard von Neumann measurement. The link between the density matrix ρ and the probabilities P(Hi) = tr(ρ Hi) is given by:

$$\rho = \sum_{i=1}^{d^2} \left[ (d+1) P(H_i) - \frac{1}{d} \right] \prod_i$$

This reconstruction is an ambitious goal that is far from being completed because it raises some very difficult mathematical problems. In particular, nobody knows today if SIC POVMs exist in general dimension (Fuchs, Hoang & Stacey 2017). This is not the place to analyse this program here and moreover, it is not necessary to understand the main conceptual points of QBism.

## 6. EPR AND NON LOCALITY

The Einstein, Podolski, Rosen (1935) paper (EPR) is probably one the most influential papers in the field of quantum foundations, and the famous criterion of reality has been discussed thousands times:

*"If, without in any way disturbing a system, we can predict with certainty (i.e., with probability equal to unity) the value of a physical quantity, then there exists an element of reality corresponding to that quantity"*.

As is well known, this argument was initially intended to show that quantum mechanics is incomplete. The authors consider a system of two particles A and B which separate. By choosing freely to measure either the position or the momentum of A, it is possible to predict with certainty either the position or the momentum of B without "disturbing" it. According to the criterion, the conclusion follows that B must have a definite position and momentum. Hence quantum mechanics is incomplete.

QBists deny this conclusion on the basis that, as we saw before, for them a probability 1 says nothing about the system itself but only means that the agent is certain that if she does a measurement she will get the expected result. So they reject the criterion of reality.

Now the standard reasoning about EPR is that either quantum mechanics is not complete and that leads to assuming hidden variables or nature is not local through what Einstein called "a spooky action at a distance". As Bell's inequalities (Bell 1964) show that local hidden variables are forbidden, the usual conclusion is that nature is not local. Indeed, since correlations between spatially separated (even by a space-like interval) measurements on A and B cannot be explained by local hidden variables, they must come from some sort of action at a distance of one measurement on the other. QBists deny this simply by noticing that (Fuchs, Mermin & Schack 2014):

*"Quantum correlations, by their very nature, refer only to time-like separated events: the acquisition of experiences by any single agent. Quantum mechanics, in the QBist interpretation, cannot assign correlations, spooky or otherwise, to space-like separated events, since they cannot be experienced by any single agent. Quantum mechanics is thus explicitly local in the QBist interpretation"*.

## 7. IS QBISM AN EPISTEMIC INTERPRETATION?

It has become common to sort interpretations of quantum mechanics according to whether they treat quantum states as corresponding directly to reality or instead represent only our knowledge of some aspect of reality. This is the now classical distinction between the ontic and epistemic interpretations of the wave function [Spekkens 2007]. The ontic view posits that quantum states refer directly to reality. The epistemic view on the contrary assumes that quantum states refer only to our knowledge of part of reality. Moreover, the quantum state can be considered as a complete description of reality - which is called the ψ-complete view - or may require to be supplemented with additional variables –this is the ψ-supplemented view [Harrigan & Spekkens 2010]. The ψ-complete view is ontic and constitutes the orthodox interpretation of quantum mechanics. However this classification relies on the assumption that there is at least an underlying ontic state. As it is not the case for QBism, it does not fit into this classification because quantum states do not represent incomplete knowledge of some underlying reality. In this sense, QBism is not an epistemic interpretation but can be called a doxastic interpretation where quantum states are states of belief of an agent. This is a point that characterizes the difference between the initial position supported by Caves, Fuchs and Schack before 2009 (Caves, Fuchs & Schack 2002, 2007) that can be called "Quantum Bayesianism" and the current QBist position. As Stacey (2019b) says quoting a letter that Fuchs sent to Mabuchi in 2000, the initial slogan "maximal information is not complete" was locked into an epistemic mindset, rather than a doxastic one.

From this point of view QBism is similar (though different in many other aspects) to some other interpretations such as Leifer & Spekkens (2013), Bub & Pitowsky (2010), Brukner & Zeilinger (2009), Brukner (2017) and to pragmatist positions (Healey 2012, 2016, 2017) which also suppose that there is no underlying ontic state.

## 8. QBISM AND THE NO-GO THEOREMS

A certain number of no-go theorems have been published recently. Pusey, Barret & Rudolph (2012) wanted to prove "*that any model in which a quantum state represents mere information about an underlying physical state of the system, and in which systems that are prepared independently have independent physical states, must make predictions which contradict those of quantum theory*". Frauchiger and Renner (2018) argued that no single-world interpretation can be logically consistent, and that quantum theory cannot be extrapolated to complex systems and in particular to observers themselves. If true these no-go theorems would be a real issue for QBism. But actually they rely on precise prior assumptions that are not correct as far as QBism is concerned (Stacey 2019a).

## 9. CRITICISMS

QBism has given rise to many criticisms since its inception. It seems that even the Wikipedia treatment of QBism has been wrong during a long time (Stacey 2016). Many criticisms were caused by earlier ambiguous or even contradictory formulations (attributed rightly or not to QBism) that have been corrected in the recent version of QBism. As Stacey (2019 b) says:

"*Fuchs and Peres present a version of the Wigner's Friend thought-experiment. In their portrayal, Erwin applies quantum mechanics to his colleague Cathy, who in turn applies it to a piece of cake. The Fuchs–Peres discussion is, from a QBist standpoint, rather unforgivably sloppy about the distinctions between ontic degrees of freedom, epistemic statements about ontic quantities and doxastic statements about future personal experiences*".

This is true when we read:

*"As soon as one detector was activated, her wave function collapsed. Of course nothing dramatic happened to her. She just acquired the knowledge of the kind of food she could eat"* (Fuchs & Peres 2000).

But, beyond these early misleading presentations, it is true that the counter-intuitive picture QBism provides has often been a source of misunderstanding, when it was not of a mere expression of reluctance to consider a position which is not compatible with standard realism. Some of these criticisms missed the point and do not need to be detailed here. See for example Nauenberg's comment (Nauenberg 2015) and the reply by Fuchs, Mermin & Schack (2015). Norsen (2014) denied that the solution given by QBism to avoid non-locality was correct since it amounts to a quantum solipsistic solution. Khrennikov (2015) made a lot of critical comments but corrected them afterwards (Khrennikov 2018). Stairs (2011) challenges the way probability-one assignments are dealt with by QBists. Mohrhoff (2014) agrees with some features of QBism such as the rejection of the EPR's reality criterion, the fact that a measurement is an act of creation, and that there is an external reality that cannot be fully represented by a quantum state. But he contests the QBist claims that a probability judgment is neither right nor wrong, that quantum correlations refer only to time-like separated events and that all external systems, whether they be atoms or other agents, are to be treated quantum-mechanically. Marchildon (2015) gives a fairly good description of the main tenets of QBism and says that, once the notion of "agent" is precisely defined (for instance, a mentally sane Homo Sapien), QBism is a consistent and well-defined theory. The definition of an agent is something on which we will come back in the next section. Marchildon's conclusion is nevertheless that he is not convinced, but he admits that it is mainly for personal preferences. Marchildon has the honesty to recognize that the picture of the world we get from QBism is so counter-intuitive that he does not like it. Many criticisms attempting to present more logical arguments against QBism are actually mere disguise for the same implicit personal feeling. Objections from Timpson (2013) are worth analysing more carefully, and we will come back to them in the next section.

## 10. DISCUSSION OF SOME ISSUES

We want to examine here questions coming neither from a bad understanding of the interpretation nor from a negative reaction due to a personal reluctance to the fact that QBism does not provide an image in line with a standard realist description of the world. Actually QBism leaves open some important questions whose answer would be necessary if one wants to get a precise detailed picture of what is going on. As Timpson (2013) says: *"Now, as a formal proposal, quantum Bayesianism is relatively clear and well developed. But it is rather less transparent philosophically."*

I analyse below a list of questions that naturally arise when one tries to go beyond a superficial understanding, which at first sight might seem satisfying but raises many issues at second sight when one tries to understand QBism more precisely. These questions do not necessarily attack the internal consistency of QBism but are intended to better understand what it would mean for one to consider it not merely an instrumentalist position, but something that aims to give a description, even if partial, of the world. Some of them will nevertheless need clear answers to be sure that QBism avoids a real inconsistency. Depending on the way these necessary clarifications are made, the global image given by QBism can vary from pure instrumentalism to interpretations making the observer play a much more radical role (as in Convivial Solipsism that we will present below) than in the usual presentation made by the founders of QBism.

The first one is to clarify what it is that makes "something" an agent. It would have been clearer for QBists to use the term "observer" which is standard in quantum mechanics instead of "agent" more often used in social science. In the context of physics, this gives rise to an ambiguity that may allow one to believe that something else than a human observer could be an agent. They argue that they do not have to define "agent":

*"No, for the same reason you don't have to in any standard decision theoretic situation. [...] Thinking of probability theory in the personalist Bayesian way, as an extension of formal logic, would one ever imagine that the motion of an agent, the user of the theory, could be derived out its conceptual apparatus?"* (DeBrota & Stacey 2019).

But this way of escaping the problem is totally illegitimate here. It is true that the standard probability theory (whether Bayesian or not) has not to define the user of the theory because the user does not play any role in the events of which probabilities quantify the occurrence. With or without any agent, the classical events occur and the agent is a passive witness to what has happened. The user is totally external. But it is not the case in QBism where, as largely emphasized by QBists, the agent is at the core of the fact that a result is gotten. It is the interaction between an agent and the external world that creates a result. Without agent, there is no result. A macroscopic apparatus is not able to create a result. So, the agent is not an entity external to the formalism as in the classical theory of probability, but an internal entity. Hence, they must define what an agent, who is necessary for creating a result, is. The way-out they use is unacceptable.

Now, according to QBists, an agent must have experience. Does that mean that an agent must be conscious or that she must have perceptions? That is something that they are reluctant to accept. They don't want to be involved in the "consciousness question". Nevertheless, they admit that experience and consciousness are difficult to disentangle (private communication). But, they say that consciousness is not necessary for QBism and that experience is all that is needed. However they admit that a computer programmed to use quantum mechanics would not be an agent and that the only agents known today are human beings… Acknowledging that an agent must be conscious to have experience would help understanding what happens when an agent has an experience and what the status of the result she gets is.

QBism is not an idealist position and accepts the existence of an external world, even if this world is not describable in usual terms. According to QBism, it is through her interaction with the external world that the agent has an experience which is the result of the measurement.

*"An experiment is an active intervention into the course of Nature: We set up this or that experiment to see how Nature reacts. We have learned something new when we can distill from the accumulated data a compact description of all that was seen and an indication of which further experiments will corroborate that description. This is what science is about. If, from such a description, we can further distill a model of a free-standing "reality" independent of our interventions, then so much the better. Classical physics is the ultimate example of such a model. However, there is no logical necessity for a realistic worldview to always be obtainable. If the world is such that we can never identify a reality independent of our experimental activity, then we must be prepared for that too."* (Fuchs & Peres 2000)

So far so good, but the following statement is much more problematic:

*"In the case of a gamble with consequences beyond an individual's expectations for their own longevity, the agent making the bet may be a community, rather than a single human being"*. (DeBrota & Stacey 2019).

Having experience is perfectly understandable for a single human being but what does it mean for a community?

The kind of realism that QBism endorses is called "participatory realism", wherein reality consists of more than can be captured by any putative third-person account of it (Fuchs 2016).

So, there seems to remain no more ambiguity about what a measurement is since there is no measurement without an agent. If one admits that an agent creates a result through her experience, it becomes acceptable to claim that the role of quantum mechanics is to give rules allowing the agent to update her old beliefs with the result she got and from that, to compute her new beliefs about future experience. But once acknowledged that the quantum formalism concerns only the modelling of the agent's belief and that quantum states, operators and Hamiltonians are subjective and relative to each agent, it remains necessary to be precise on what, independently of the formalism, bears her belief. What is the ontology? QBists are not very clear on it. Timpson (2013) tries painfully to give a description and warns the reader that, perhaps, the founders of QBism will not endorse it. The simple fact that it is necessary for Timpson to guess the ontology of QBism without even being sure that his guess will be faithful to what QBists have in mind, shows that there is a genuine obscurity on this point. QBism endorses the existence of an external world independent of any agent, but it is not clear if the external world is unique and shared by all agents or if each agent has her own external world.

*"... the real world, the one both are embedded in –with its objects and events – is taken for granted. What is not taken for granted is each agent's access to the part of it he has not touched"* (Fuchs 2010a, 2010b).

QBism claims that the micro level of this world is something unspeakable. There is no law ruling the behaviour of micro-objects. Any event that occurs when a system and a measuring device interact is not determined by anything, not even probabilistically: *"Something new really does come into the world when two bits of it [system and apparatus] are united"* (Fuchs 2006). Two questions then arise. The first one is: how does this result appear from the interaction between a system at the micro-level which is unspeakable and without law, and a measuring device? Timpson proposes that systems have dispositions to give rise to events when they interact with one another. He says that we can think of the world at each time as pregnant with possibilities, while yet, what will happen is completely undetermined:

*"The micro-objects are seats of causal powers and there is nothing to explain why they give rise to the particular manifestations of their powers that they do"*. (Timpson 2013).

This is reminiscent of the "opium dormitive virtue" that Molière ridiculed. It is the kind of explanation that explains nothing and, I think it would be more acceptable to say simply that we cannot talk of the process of giving rise to an event. The second question is the status of the result the agent gets. I think that it would be unfaithful to QBism to say that the result appears as a simple interaction between a quantum system and an apparatus and that the agent simply records it. It is probable that QBists would not agree today with Fuchs' sentence above where he said that something new does come into the world when a system and an apparatus are united, since it does not make any mention of an agent in the measurement process. That would come back to the Copenhaguen interpretation and this is not what QBism says. A result is something happening in the agent's experience and without the agent, there is no result. QBism says on the one hand that the external world is independent of any agent and that a result comes from the interaction between an agent and the external world, and on the other hand that the result is only for the agent herself:

*"It says that the outcomes of quantum measurements are single-user as well! That is to say, when an agent writes down her degrees of belief for the outcomes of a quantum measurement, what she is writing down are her degrees of belief about her potential personal experiences arising in consequence of her action on the external world."* (Fuchs & Schack 2014).

So the question is: is the result in the agent's mind or is it in the external world, which would imply that each agent owns her personal external world? If it is only in the mind of the agent, then the result is non-physical, and it is not true that "something new comes into the world". If the result is in the external world, how is it possible that the agent creates this result (which is not created simply by the interaction between the system and the apparatus)? It seems that we are back to the position of London and Bauer (1939) with a mental process that can have a physical impact on the world while resting outside of the scope of quantum mechanics. And if the result is in the agent's personal external world, is it shared with other agents or is it only available for the agent having created it? According to Fuchs & Schack (2014):

*"But by that turn, the consequences of our actions on physical systems must be egocentric as well. Measurement outcomes come about for the agent herself. Quantum mechanics is a single-user theory through and through – first in the usual Bayesian sense with regard to personal beliefs, and second in that quantum measurement outcomes are wholly personal experiences."*

So it seems that the result of a measurement made by an agent is not necessarily shared by another agent. But in this case, is this result a modification of something in the external world or not?

*"One often think of events as involving the modification of properties of objects, so don't we need some occurrent properties of the systems to be around to be modified if there are to be any events? Perhaps. But again we can (and should) be agnostic on the details."* (Timpson 2013).

The fuzziness of that sentence clearly shows that many questions remain obscure. As we will see, these questions find natural answers in Convivial Solipsism.

A second issue is related to the troubles QBism seems to have with explanation. If QBism considers that the quantum formalism concerns exclusively agents' beliefs then, physics does not speak of the world. As Timpson (2013) says *"Ultimately we are just not interested in agents' expectation that matter structured like sodium would conduct; we are interested in why in fact it does so"*. QBists' answers, relying on historical examples like Kepler's wrong explanation of the number of planets or on the abandon of the phlogiston (DeBrota & Stacey 2019) to clear one's name, are not really convincing. It is true that the status of what is a good explanation has strongly changed from the Greeks to our time, but that does not mean that we don't feel when something provides no explanation at all, and that seems to be the case with QBism, especially with its emphasis about the indescribable aspect of the external world and the total absence of any law.

Now, there are some issues coming from the QBists' claim that all probabilities are purely subjective in the quantum framework. De Finetti, Ramsey and Savage had not in mind quantum mechanics when they designed their position. The subjective concept of probabilities makes sense in a deterministic world where the subjectivity comes from our ignorance of the details allowing in principle to know the future result. But QBism does not say that quantum mechanics is ultimately deterministic. On the contrary, it even explicitly says that the result an agent gets during a measurement is fundamentally at random, and that there is an objective indeterminism. This makes a big difference with the context De Finetti, Ramsey and Savage had in mind. So, it is difficult to deny that quantum probabilities, linked with the occurring of an objectively random event do not contain, at least a small, objective part. Mentioning a micro-level which is unspeakable, without law, and saying that there are no facts that determine the measurement outcome is not enough to support the claim that all probabilities are subjective. On the contrary, this seems to prove that something fundamentally random is at work and that objective probabilities are there, even if we cannot know them. Objective probabilities do not need to be computable to exist. QBism can coherently claim that these objective probabilities are not knowable, that they are out of the scope of the quantum formalism and that all probabilities taken into account in the formalism are subjective, but it cannot claim that only subjective probabilities exist. Moreover, the problem of explaining why the subjective probabilities provided by the formalism are better than anything else remains open. One answer could be that these subjective probabilities are based on an attempt to guess the underlying objective probabilities. But QBists refuse this answer and adopt an even more radical position. After a measurement having given the result d, the post measurement state is updated to |d> but that does not mean that the system is in the state |d> since this assignation only concerns the agent's state of belief. Of course, the probability for the agent to get the result d if she performs again the same measurement is 1, so the agent is sure to get d. But as in QBism probability 1 does not imply anything objective for the system, it does not need to be "really" in the state |d>. The question is then: since the objective existence of the system is admitted by QBists, what can be said about the system? We come back to the questions we raised above. Has the system changed? If the answer is no, then what does it mean that the agent has created a non-pre-existing result in her interaction with the system? Is the result she saw only in her mind? And if the answer is yes, why not consider that the new state |d> objectively represents something (perhaps incomplete) about the system, be it only true for the agent herself? As Timpson (2013) says:

*"Of course, this does not mean that whether or not an apparatus provides a particular read-out or any read-out at all, is a subjective matter; on measurement one of the outcomes d will, objectively, obtain."*

So there is a link between the objective result d and the update of belief. Not recognizing that and maintaining at all cost that the state only representis the agent's subjective degree of belief seems a little bit artificial. That means that an agent can be certain that some particular outcome will obtain and simultaneously claim that there is no fact about what that outcome will be. That is fundamental for preserving the status QBism gives to probability 1, which is in turn necessary for avoiding the EPR criterion of reality. As Timpson (2013) rightly notices, this amounts to what he calls the "Quantum Bayesian Moore's paradox": an agent is committed to say at the same time "I am certain that p and it is not certain that p". This is certainly problematic and the solution Timpson charitably proposes is not very convincing:

*"The two halves of the pertinent sentences concern different levels of discussion; that the second claim – 'it is not certain' – is made at a meta level of philosophical analysis; is said, perhaps, in a different tone of voice, from the ground level – 'I am certain'-; and thus the tension is diffused."*

That seems more like a joke than a real solution …

At this stage, I do not claim that there is any definitive contradiction in QBism but only that all these points should be made clear in order to be able to get a global coherent picture. All the questions above may well find answers if things are made clearer about the status of the external world and what it means to be the result that an observer gets from a measurement. These issues find a natural explanation in Convivial Solipsism which is much more explicit on all these points. In the next paragraph, I will show that answering these questions in a natural way and diminishing the emphasis put on the subjective probabilities, would lead QBism to be very near to Convivial Solipsism.

### 11. COMPARISON WITH CONVIVIAL SOLIPSIM (ConSol)

Convivial Solipsism (ConSol), which in spite of his name is not a solipsist position but a kind of realist one, has been exposed first in a French book (Zwirn 2000) and further developed in (Zwirn 2015, 2016, 2017, 2019). It starts from the desire to get rid of the physical collapse of the wave function exactly as in Everett's interpretation (Everett 1956). But the initial motivation for Convivial Solipsism comes from a remark of d'Espagnat (1971) who was totally unsatisfied with the astronomical number of worlds (or of minds) that is assumed in the various presentations of Everett's interpretation. ConSol is an interpretation inside which the physical dynamics of the universe is described by the Schrödinger equation and which states that a measurement is nothing else than the awareness of a result by a conscious observer. ConSol assumes that there is an external world that each observer models with her own wave-function. This external world is not directly describable in terms of classical properties. Contrary to QBism, the wave function is considered as modelling (in a very limited way) the external world but a system represented by such a superposed wave function cannot be perceived by a human observer in all its richness. Consequently, an observer watching a superposed physical thing will perceive it through some mental filters allowing her to see only a definite result and giving her the impression that a definite result is "really" obtained and that the system is "really" in the state she perceives.

*"Convivial Solipsism is situated in a neo-Kantian framework and assumes that there is "something" else than consciousness, something that (according to the famous Wittgenstein's sentence) it is not appropriate to talk of. This is close to what Kant calls "thing in itself" or "noumenal world". Consciousness and this "something" give rise to what each observer thinks is her reality, following Putnam's famous statement "the mind and the world jointly make up the mind and the world". So perception is not a passive affair: perceiving is not simply witnessing what is in front of us but is creating (independently for each us) what we perceive through a co-construction from the world and the mind. The hanging-on mechanism takes part in this co-construction and helps (very partially) understanding it through the selection it does."[2]* (Zwirn 2016).

---

[2] Of course, it is out of question to be able to explain how awareness happens. This is probably one of the most difficult problems of contemporary science. Neuroscientists are hardly beginning to understand some mechanisms that show how consciousness works and how different it is from what our own consciousness itself thinks it is working.

In QBism probabilities are agent's degrees of belief about the result she will get if she performs an experiment. ConSol puts less emphasis on the subjective aspect of probabilities, which is not the important point. Probabilities are simply about the observer's states of awareness because between many possible ones, only one appears. Indeed actually, what she sees is just the result of the limitation of her perception by these filters, and the system remains in a superposed state. This awareness has no physical impact on the systems that are measured and whose states are unchanged after the measurement. The reduction is only a way to describe what appears to the observer and does not affect the external world which remains superposed. Hence there is no physical collapse but contrary to what happens in Everett's interpretation, the world does not split in as many versions as there are possible results. If we compare this with QBism, we see that the result of a measurement is also created in the interaction of the observer with the external world. Unlike QBism, ConSol relies on the decoherence mechanism to solve the problem of the preferred basis (Zwirn 1997, 2015, 2016). But a major common point between QBism and ConSol is that both say that quantum mechanics is about the interaction of an agent or an observer and an external world knowable at best very partially and imperfectly. But ConSol is clear about the fact that this result exists only in the observer's perception, and the reason why an observer gets a unique result is clear: that comes from the limited perceptive capacity of the observer who is unable to see a superposed state in its whole richness. QBism and ConSol both claim that it is not possible to give any familiar complete description of the external world:

*"Perhaps it is just the case that once one seeks to go beyond a certain level of detail, the world simply does not admit of any straightforward description or capturing by theory, and so our best attempts at providing such a theory do not deliver us with what we had anticipated, or with what we wanted. [...] The world, perhaps, to borrow Bell's felicitous phrase, is unspeakable below a certain level."* (Timpson 2013).

ConSol rests on two main assumptions. The first one is the hanging-on mechanism which describes what a measurement is and how an observer must update her wave-function after getting a result:

*"A measurement is the awareness of a result by a conscious observer whose consciousness selects at random (according to the Born rule) one branch of the entangled state vector written in the preferred basis and hangs-on to it. Once the consciousness is hung-on to one branch, it will hang-on only to branches that are daughters of this branch for all the following observations."* (Zwirn 2016).

The consequence is that even if the external world is not impacted by the measurement, everything goes for the observer as if a real result has been created and she must update her wave-function by restricting it to the branch corresponding to the result she got. There is both a similarity and a difference in the way QBism and ConSol deal with measurements. The similarity is that a measurement is necessarily an interaction between an agent or an observer and the external world, and that a measurement is an act of creation. It is not a simple record of a pre-existing state of affairs. The difference is that in QBism, it seems (but that is not so clear) that the result is "really created" in the external world of the agent while in ConSol, nothing changes in the external world, the result being entirely inside the observer's perception. As QBism, ConSol assumes that the quantum formalism is universal and can be used to model not only macroscopic systems but also other observers. That means that an observer can use a description of another observer through a superposed state encoding her probabilities for the possible answers to any question she might ask before getting a definite answer. The hanging-on mechanism guarantees that two observers will never disagree on the result of a measurement because any communication between the two is like a measurement of the one by the other (and vice versa). This is the reason why it is called Convivial. Another similarity between QBism and ConSol is that quantum states (and Hamiltonians and operators) are considered relative to the observer. QBism is clear about the personal use of quantum mechanics but is fuzzy about when a result created by an agent is shared with other agents. In ConSol, this question has no meaning because it is forbidden to speak simultaneously of the perceptions of two observers. Each sentence about a result has to be stated only from the point of view of a unique observer. Sentences like "Alice saw 'up' and Bob saw 'down'" are not allowed. It seems that it is also the case for QBism since otherwise, QBism's solution to non-locality would not work. Unlike QBism, ConSol does not consider that probability 1 does not give any information on the system, but nevertheless it avoids non-locality for the same reason as QBism: no single observer can make space-like separated measurements. Hence, if QBists are ready to go that far to accept the fact that it is impossible to compare the results gotten by two agents, the picture given by QBism and ConSol would be similar regarding the fact that "what is real" is only valid for one observer and that a third-person point of view is forbidden. Each observer lives in her world and QBism would merit also to be called "a kind of solipsism" (in the weak sense adopted in ConSol) even though this is something QBists hate.

## 12. CONCLUSION

Many criticisms against QBism or ConSol come from the same source:

*"Other confusions typically occur when the interlocutor has unwittingly switched from subjective probability to objective, or from a first-person perspective to third-person, midway through a thought process. These are habits which take discipline to avoid, at least at first"* (Stacey 2019b).

In the case of ConSol, the confusion does not come from switching from subjective probability to objective, but from switching from a first-person perspective to third person. These two interpretations give a picture of the world which is radically different from what the majority of physicists are accustomed to, and even are ready to accept. In the list of the many interpretations that have been proposed, it is first necessary to eliminate those which are inconsistent (and there are many of them). Then the choice is largely a matter of taste since, at the moment, no reasonable and doable experiments exist to decide between those remaining.

**Acknowledgment**

I want to thank Chris Fuchs for the vivid exchanges we have had in the colloquium at "Les Treilles" and in Paris, and through many emails allowing me to better understand QBism. I want also to thank Ruediger Schack and David Mermin for former enlightening exchanges.